\documentclass[journal]{IEEEtran}
\usepackage[T1]{fontenc}
\usepackage{psfrag,amsmath,amssymb,color,cite, amsmath,graphicx}
\usepackage{stfloats}
\usepackage[caption=false,font=footnotesize]{subfig}
\usepackage{algorithm}
\usepackage{mathrsfs}
\usepackage[algo2e,ruled,vlined]{algorithm2e}
\usepackage{algpseudocode}
\usepackage{pifont}
\usepackage[colorinlistoftodos]{todonotes}

\newtheorem{remark}{Remark}

\newcommand{\beqn}{\begin{equation}}
\newcommand{\eeqn}{\end{equation}}
\newcommand{\beqa}{\begin{eqnarray}}
\newcommand{\eeqa}{\end{eqnarray}}
\newcommand{\beqas}{\begin{eqnarray*}}
\newcommand{\eeqas}{\end{eqnarray*}}

\DeclareMathOperator*{\argmax}{arg\,max}

\presetkeys{todonotes}{color=green!30}{}
\newcommand{\vz}{{{\bf z}} }
\newcommand{\vy}{ {{\bf y}} }
\newcommand{\vx}{ {{\bf x}} }
\newcommand{\mh}{ {{\bf H}} }

\begin{document}

\title{Low-Complexity Iterative Detection for \\ Orthogonal Time Frequency Space Modulation}
\author{\IEEEauthorblockN{P. Raviteja, Khoa T. Phan, Qianyu Jin, Yi Hong, and Emanuele Viterbo\\}
\IEEEauthorblockA{ECSE Department, Monash University, Clayton, VIC 3800, Australia\\
Email: \{raviteja.patchava, khoa.phan, qianyu.jin, yi.hong, emanuele.viterbo\}@monash.edu}}
\maketitle

\begin{abstract}
We elaborate on the recently proposed orthogonal time frequency space (OTFS) modulation technique, which provides significant advantages over orthogonal frequency division multiplexing (OFDM) in
Doppler channels. We first derive the input--output  relation describing OTFS modulation and demodulation (mod/demod) for delay--Doppler channels with arbitrary number of paths, with given delay and Doppler values. We then propose a low-complexity message passing (MP) detection algorithm, which is suitable for large-scale OTFS taking advantage of the inherent channel sparsity.
Since the fractional Doppler paths (i.e., not exactly aligned with the Doppler taps) produce the inter Doppler interference (IDI), we adapt the MP detection algorithm to compensate for the effect of IDI in order to further improve performance.
Simulations results illustrate the superior performance gains of OTFS over OFDM under various channel conditions.
\end{abstract}
\section{Introduction}
Fifth-generation (5G) mobile systems are expected to accommodate an enormous number of emerging wireless applications with high data rate requirements (e.g., real-time video streaming, and online gaming, connected and autonomous vehicles etc.). While the orthogonal frequency division multiplexing (OFDM) modulation scheme currently deployed in fourth-generation (4G) mobile systems achieves high spectral efficiency for time-invariant frequency selective channels, it is not robust to time-varying channels, especially for channels with high Doppler spread (e.g., high-speed railway mobile communications). Hence, new modulation schemes/waveforms that are robust to channel time-variations are being extensively explored.

Recently, orthogonal time frequency space
(OTFS) modulation was proposed in \cite{Hadani} showing  significant advantages over OFDM, in
delay--Doppler channels with a number of paths, with given delay and Doppler values. The delay-Doppler domain is an alternative representation of a time-varying channel geometry due to moving objects (e.g. transmitters, receivers, reflectors) in the scene.  Leveraging on this representation, the OTFS modulator spreads each information (e.g., QAM) symbol over a set of two dimensional (2D) orthogonal basis functions, which span across the frequency--time resources required to transmit a burst.  The  basis function set is specifically designed to combat the dynamics of the
time-varying multi-path channel. The general framework of OTFS was given in \cite{Hadani} and a {\it coded} OTFS system with forward error correction (FEC) and turbo equalization was compared with coded OFDM, showing significant gain.

In this paper, we analyze the input-output relation describing {\em uncoded} OTFS modulation/demodulation for delay--Doppler channels with a number of paths, with given delay and Doppler values. We then propose a low-complexity message passing (MP) detection algorithm, which is suitable for large-scale uncoded OTFS taking advantage of the inherent channel sparsity.
The MP detection algorithm, based on a sparse factor graph, uses Gaussian approximation of the interference terms to further reduce the complexity, similar to \cite{mp_ga} which was applied to massive MIMO detection. Since the fractional Doppler paths (i.e., not exactly aligned with the Doppler taps) produce the inter Doppler interference (IDI), we adapt the MP detection algorithm to compensate for the effect of IDI in order to further improve performance.
We show that the proposed MP detection algorithm can also be applied to OFDM systems to compensate for the Doppler effects. Through simulations, we show the superior performance gains of OTFS over OFDM under various channel conditions.





\begin{figure*}
\centering
\includegraphics[width=7in]{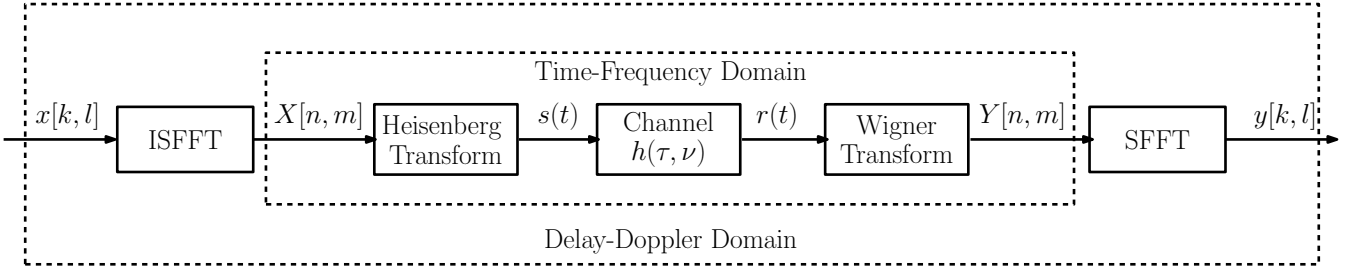}
\caption{OTFS mod/demod}
\vspace{-5.5mm}
\label{sys_fig}
\end{figure*}

\section{OTFS modulation/demodulation}
In this section, we describe OTFS modulation/demodulation \cite{Hadani}.

\subsection{General OTFS block diagram}
The OTFS system diagram is given in Fig. \ref{sys_fig}. OTFS modulation is produced by a cascade of two 2D transforms at both the transmitter and the receiver.  The modulator first maps the information symbols $x[k,l]$ in the delay--Doppler domain to symbols $X[n,m]$ in the time--frequency domain using {\em inverse symplectic finite Fourier transform} (ISFFT). Next, the {\em Heisenberg transform} is applied to time--frequency symbols $X[n,m]$ to create the time domain signal $s(t)$ transmitted over the wireless channel.
At the receiver, the received time-domain signal $r(t)$ is mapped to the time--frequency domain through the {\em Wigner transform} (the inverse of the Heisenberg transform), and then to the delay--Doppler domain for symbol demodulation.

We first introduce the following notation:
\begin{itemize}
\item The time-frequency plane is discretized by sampling time and frequency axes at intervals $T$ (seconds) and $\Delta f$ (Hz), respectively:
\beqn
\Lambda = \bigl\{(nT,m\Delta f),\;\; n=0,\hdots,N-1, m=0,\hdots,M-1\bigr\}. \nonumber
\eeqn
\item A packet burst has duration $NT$ and occupies a bandwidth $M\Delta f$.
\item Modulated symbol set $X[n,m], n=0,\hdots,N-1, m=0,\hdots,M-1$ is transmitted over a given packet burst.
\item Transmit and receive pulses are denoted by $g_{\text{tx}}(t)$ and $g_{\text{rx}}(t)$.
\end{itemize}
Moreover, the delay--Doppler plane is discretized as follows:
\beqn
\Gamma = \Bigl\{\left(\frac{k}{NT},\frac{l}{M\Delta f}\right),\; k=0,\hdots,N-1, l=0,\hdots,M-1\Bigr\}, \nonumber
\eeqn
where $\frac{1}{NT}$ and $\frac{1}{M\Delta f}$ represent the quantization intervals of the Doppler frequency shift and time delay, respectively.
\subsection{OTFS modulation}
 Consider a set of $N \times M$ information symbols $x[k,l], k=0,\hdots,N-1, l=0,\hdots, M-1$ from a modulation alphabet $\mathbb{A} = \{ a_1, \cdots, a_{|\mathbb{A}|} \}$ (e.g. QAM), which are arranged on a 2D delay--Doppler grid $\Gamma$ that we wish to transmit.

The OTFS maps $x[k,l]$ to symbols $X[n,m]$ in the time--frequency domain using inverse symplectic finite Fourier transform (SFFT) as follows:
\beqn
X[n,m] = \frac{1}{NM}\sum_{k=0}^{N-1}\sum_{l=0}^{M-1} x[k,l] e^{j2\pi\bigl(\frac{nk}{N}-\frac{ml}{M}\bigr)} \label{iSFFT}
\eeqn
for $n=0,\hdots,N-1, m=0,\hdots,M-1$.

Next, a time--frequency modulator maps $X[n,m]$  on the grid $\Lambda$ to a transmitted waveform $s(t)$ by
\beqn
s(t) = \sum_{n=0}^{N-1}\sum_{m = 0}^{M-1}   X[n,m] g_{\text{tx}}(t-nT) e^{j2\pi m \Delta f(t-nT)}. \label{otfs1}
\eeqn
\begin{remark}  
The modulation operation in (\ref{otfs1}) generalizes OFDM which maps information symbols from frequency domain to time domain.
If $g_{\text{tx}}(t)$ is a rectangle waveform with duration of $T$ then (\ref{otfs1}) reduces to conventional inverse discrete Fourier transform. When $N=1$, the inner box in Fig. \ref{sys_fig} is an OFDM system. Therefore, one OTFS symbol (packet burst) can be viewed as a SFFT precoding applied on $N$ consecutive independent OFDM symbols with $M$ subcarriers.
\end{remark}
\subsection{Wireless transmission and reception}
The signal $s(t)$ is transmitted over a time-varying channel with complex baseband channel impulse response $h(\tau,\nu)$,
which characterizes the channel response to an impulse with delay $\tau$ and Doppler $\nu$ \cite{Jakes}. The received signal $r(t)$ is given by:
\beqn
r(t) = \int \int h(\tau,\nu) s(t-\tau) e^{j2\pi\nu(t-\tau)}d\nu d\tau + v(t) \label{delay-doppler}
\eeqn
where $v(t)$ is the additive noise at the receiver.


\subsection{OTFS demodulation}
\subsubsection{Sufficient statistics and channel distortion}
The matched filter computes the cross-ambiguity function $A_{g_{\text{rx}},r}(t,f)$:
\beqn
Y(t,f) = A_{g_{\text{rx}},r}(t,f) \triangleq \int g_{\text{rx}}^*(t'-t) r(t') e^{-j 2 \pi f (t'-t)} dt'. \label{wigner1}
\eeqn
The matched filter output  can be obtained by sampling the function $Y(t,f)$ at $t = nT$ and at $f = m \Delta f$:
\beqn
Y[n,m] = Y(t,f)|_{t = nT, f = m \Delta f}. \label{wigner}
\eeqn
The operation of (\ref{wigner1}) and (\ref{wigner}) is  called  {\em Wigner transform}.
\begin{remark}
Wigner transform is a generalization of the OFDM receiver, which maps the received time domain signal to the frequency domain modulated symbols.
When $g_{\text{rx}}(t)$ is rectangle waveform, it corresponds to the discrete Fourier Transform in  OFDM.
\end{remark}

The relationship between the matched filter output $Y[n,m]$ and the transmitter input $X[n,m]$ was established in \cite{Hadani} as:
\beqn \label{eq:receivedsig_Y}
Y[n,m] = H[n,m]X[n,m] + V[n,m]
\eeqn
where $V(t,f) = A_{g_{\text{rx}},v}(t,f) $, and $V[n,m] = V(t,f)|_{t = nT, f = m \Delta f}$, and
\beqn
H[n,m] = \int \int h(\tau,\nu) e^{j 2\pi \tau nT}e^{-j 2\pi(\nu + m\Delta f)\tau}d\nu d\tau \nonumber
\eeqn
given that $h(\tau,\nu)$ has finite support bounded by $(\tau_{\rm max},\nu_{\rm max})$, and the pulses $g_{\text{tx}}(t)$ and $g_{\text{rx}}(t)$ are ideal, i.e., they satisfy the condition $A_{g_{\text{rx}}, g_{\text{tx}}}(t,f) = 0$ for $t \in (nT-\tau_{\rm max},nT+\tau_{\rm max})$, $f \in (m\Delta f-\nu_{\rm max},m\Delta f+\nu_{\rm max})$, where $\tau_{\rm max}$, and $\nu_{\rm max}$ are the maximum delay and Doppler values among channel paths. The condition  on ideal pulses is called the {\em bi-orthogonal property} and it does not hold for practical pulses (for example, rectangular pulses). Nonetheless, we assume ideal pulses as in \cite{Hadani}, and the practical pulses are discussed in \cite{fut_pap}.


Next we apply the SFFT on $Y[n,m]$, yielding \cite{Hadani}:
\begin{align}
& y[k,l] = \frac{1}{NM}\sum_{n=0}^{N-1} \sum_{m=0}^{M-1} Y[n,m] e^{-j2\pi\bigl(\frac{nk}{N}-\frac{ml}{M}\bigr)} \nonumber \\
& =  \frac{1}{NM}\! \sum_{n=0}^{N-1} \sum_{m=0}^{M-1} x[n,m] h_w\Bigl(\frac{k-n}{NT}, \frac{l-m}{M\Delta f}\Bigr) \! + \! z [k,l],
\label{rx_symbol}
\end{align}
where $z [k,l] \sim \mathcal{CN} (0, \sigma^2)$ is the additive white noise, and $h_w(\cdot,\cdot)$ is a sampled version of the impulse response function:
\beqn
h_w\Bigl(\frac{k-n}{NT}, \frac{l-m}{M\Delta f}\Bigr) = h_w(\nu',\tau')|_{\nu'=\frac{k-n}{NT},\tau'=\frac{l-m}{M\Delta f}} \nonumber
\eeqn
for $h_w(\nu',\tau')$ being the circular convolution of the channel response with a windowing function: 
\beqa
h_w(\nu',\tau') \!\!&=&\!\! \int \int h(\tau,\nu) w(\nu' - \nu, \tau' -\tau) e^{-j2\pi\nu\tau}d\tau d\nu, \nonumber\\
w(\nu,\tau) \!\!&=\!\!& \sum_{c=0}^{N-1} \sum_{d=0}^{M-1} e^{-j 2\pi (\nu cT - \tau d\Delta f)}. \nonumber
\eeqa
Here, we assume that the rectangular window is applied on the transmitter and receiver symbols $X[n,m]$ and $Y[n,m]$.


\section{OTFS under sparse delay--Doppler channel representation}
\subsection{Sparse representation of the delay--Doppler channel}
A sparse representation of the delay--Doppler channel $h(\tau,\nu)$ in (\ref{delay-doppler}) can be expressed as:
\begin{equation}
h(\tau,\nu) = \sum_{i=1}^{P} h_i \delta(\tau-\tau_i) \delta(\nu-\nu_i)
\label{del_dop}
\end{equation}
where $P$ is the number of reflectors; $h_i$, $\tau_i$, and $\nu_i$ represent the channel gain, delay, and Doppler shift associated with $i^{th}$ reflector, respectively. Here, we assume that the delays for each reflector are different.

We now analyze the received symbols $y[k,l]$ in (\ref{rx_symbol}) using the sparse presentation of the delay--Doppler channel $h(\tau,\nu)$ in (\ref{del_dop}).

Let us consider the expression for $h_w(\nu',\tau')$ by substituting the delay--Doppler channel  in (\ref{del_dop}),
\begin{align}
& h_w(\nu',\tau')  = \sum_{i=1}^{P} h_i e^{-j2\pi\nu_i\tau_i} \, w(\nu' - \nu_i, \tau' -\tau_i)
\nonumber \\
& = \sum_{i=1}^{P} h_i e^{-j2\pi\nu_i\tau_i} \sum_{c=0}^{N-1} e^{-j 2\pi (\nu' - \nu_i) cT} \sum_{d=0}^{M-1} e^{j 2\pi (\tau' -\tau_i) d\Delta f} \nonumber \\
& = \sum_{i=1}^{P} \, h'_i \, \mathcal {G} (\nu',\nu_i) \, \mathcal {F} (\tau',\tau_i),
\end{align}
where $h'_i = h_i e^{-j2\pi\nu_i\tau_i}$, $\mathcal {F} (\tau',\tau_i) = \sum_{d=0}^{M-1} e^{j 2\pi (\tau' -\tau_i) d\Delta f}$, and $\mathcal {G} (\nu',\nu_i) = \sum_{c=0}^{N-1} e^{-j 2\pi (\nu' - \nu_i) cT}$.

Let us first evaluate $\mathcal {F} (\tau',\tau_i)$ at $\tau' = \frac{l-m}{M\Delta f}$,
\begin{align}
\mathcal {F} \, \left(\frac{l-m}{M\Delta f},\tau_i\right) = \sum_{d=0}^{M-1} e^{j \frac{2\pi}{M} (l - m - \alpha_i) d}= \frac{e^{j {2\pi} (l - m - \alpha_i) }-1}{e^{j \frac{2\pi}{M} (l - m - \alpha_i)}-1}.
\label{f_eq}
\end{align}
Here, we assume that $\tau_i= \frac{\alpha_i}{M\Delta f}$, where $\alpha_i$ is a positive integer, as the received signal is sampled at interval $\frac{1}{M\Delta f}$ \cite{wc_book}. From (\ref{f_eq}), we see that
\[
\mathcal {F} \, \left(\frac{l-m}{M\Delta f},\tau_i\right) = \left.
 \begin{cases}
 M, & \mbox{if } [l-m-\alpha_i]_M = 0 \\
 0, & \mbox{otherwise }
 \end{cases}
 \right.,
\]
where $[\cdot]_M$ represents mod $M$ operation. Hence, the function $\mathcal {F} \, \left(\frac{l-m}{M\Delta f},\tau_i\right)$ evaluates to $M$ only if $m = [l - \alpha_i]_M$, and $m \in \{0,\cdots,M-1\}$.

Similarly, we have
\begin{align}
\mathcal {G} \, \left(\frac{k-n}{NT},\nu_i \right) & = \frac{e^{j {2\pi} (k - n - \beta_i - \gamma_i) }-1}{e^{j \frac{2\pi}{N} (k - n - \beta_i- \gamma_i)}-1} .
\label{g_val}
\end{align}
Here, we assume $\nu_i = \frac{(\beta_i+\gamma_i) }{NT}$, with an integer $\beta_i$ and $0 < \gamma_i < 1$ (i.e., Doppler frequencies are not necessarily at the sampling points in the delay-Doppler plane). Specifically, $\alpha_i$ and $\beta_i$ represents the indexes of the delay tap and Doppler frequency tap,  corresponding to delay $\tau_i$ and Doppler frequency $\nu_i$, respectively.
The negative indexes of the Doppler frequency taps, where $\beta_i<0$, can also be view as those of positive frequency taps considering mod $N$ operation.

We recall that the delay--Doppler channel can be seen as a $N\times M$ discretized grid $\Gamma$ with $N$ and $M$ representing the indexes of the maximum Doppler and delay taps, respectively.
We will refer to $\gamma_i$ as the {\em fractional Doppler} since it represents the fractional shift from a Doppler tap $\beta_i$ in $\Gamma$.

The magnitude of the function $\frac{1}{N} \mathcal {G} \left(\frac{k-n}{NT},\nu_i \right)$ is
\begin{align}
\left|\frac{1}{N} \mathcal {G} \left(\frac{k-n}{NT},\nu_i \right)\right| & = \left|\frac{\sin \left( \pi \left( (k-n-\beta_i) - \gamma_i \right) \right)} {N \sin \left( \frac{\pi}{N} \left( (k-n-\beta_i) - \gamma_i \right) \right)} \right| \nonumber \\
& \geq \left| {\mbox{sinc}} \left( \pi \left( (k-n-\beta_i) - \gamma_i \right) \right) \right|.
\label{sinc}
\end{align}
The above lower bound is tight for small values of $\frac{\pi}{N} \left( (k-n-\beta_i) - \gamma_i \right)$. When $\gamma_i = 0$, the above function has the peak of the main lobe at $n = k - \beta_i$ and the peaks of the side lobes decay at rate of $1/\pi \left( k-n-\beta_i \right)$.

Therefore, for a given $k$ and $\beta_i$, the function of $\gamma_i$ in (\ref{sinc}) has the following properties:
\begin{enumerate}
\item Two-sided decreasing function with the peak at $n = [k - \beta_i]_N$, for $0 < \gamma_i \leq 0.5$.
\item Two-sided decreasing function with the peak at $n = [k - \beta_i + 1]_N$, for $0.5 < \gamma_i < 1$.
\end{enumerate} 
Therefore, we only consider a finite number ($2E_i +1$) of significant values of $\mathcal {G} \left(\frac{k-n}{NT},\nu_i \right)$ in (\ref{g_val}) for $n = [k - \beta_i + q]_N$ and $-E_i \leq q \leq E_i$, where $E_i\ll N$ (e.g., $E_i = 5$ for $N = 128$).

Then we can rewrite the receive signal $y[k,l]$ in (\ref{rx_symbol}) as
\begin{align}
y[k,l] = & \sum_{i=1}^{P} \sum_{q=-E_i}^{E_i}  h'_i  \left(\frac{e^{j {2\pi} (- q - \gamma_i) }-1}{N e^{j \frac{2\pi}{N} (- q - \gamma_i)}-1}\right) \cdot \nonumber \\
& x\left[[k - \beta_i + q]_N, [l - \alpha_i]_M\right] + z [k,l].
\label{conv_eq}
\end{align}

The above input-output expression simplifies for the following special cases.

{\em i) Ideal channel} -- Assuming $h(\tau,\nu) = \delta(\tau) \delta(\nu)$, the received signal becomes
\begin{align}
y[k,l] & = x[k,l] + z [k,l],
\nonumber
\end{align}
and behaves as an AWGN channel.

{\em ii) No fractional Doppler} $(\gamma_i=0$ for $i=1,\ldots, P)$ -- Assuming that Doppler frequencies are the exact integer multiples of Doppler taps, the received signal can be obtained by replacing $E_i = 0$ in (\ref{conv_eq}), i.e.,
\begin{align}
y[k,l] = \sum_{i=1}^{P} h'_i x[[k - \beta_i]_N, [l - \alpha_i]_M] + z [k,l].
\nonumber
\end{align}
For each path, the transmitted signal is circularly shifted by the delay and Doppler taps and scaled by the associated channel gain.

From (\ref{conv_eq}), we can see that with the fractional Doppler, the transmitted signal not only shifts by the delay and Doppler taps but also affects the neighboring Doppler taps ($-E_i$ to $E_i$). We refer to this interference on the neighboring Doppler taps as {\em inter Doppler interference (IDI)}.

\section{Message passing detection algorithm for OTFS}
We now propose a message passing (MP) detection algorithm for OTFS using the input-output relation in (\ref{conv_eq}).
\subsection{Low-complexity MP detection algorithm for OTFS}
The received signal in vectorized form can be written as
\begin{align}
\vy = \mh \, \vx + \vz,
\label{vec_form}
\end{align}
where $\vy \in \mathbb{C}^{NM \times 1}, \mh =\{{H}_{d,c}\}\in \mathbb{C}^{NM \times NM},$ and $\vx \in \mathbb{C}^{NM \times 1}$. The $(k+N l)$-th element of $\vy$ is ${y}_{k+N l} = y[k,l],$ for $k =0 ,\cdots,N-1$, $l = 0,\cdots, M-1$.
The elements of $\vx$ and $\vz$ are similarly related to $x[k,l]$ and $z[k,l]$, respectively.
Due to mod $N$ and mod $M$ operations in (\ref{conv_eq}),
we observe that only ${S} = \sum_{i=1}^P (2E_i+1)$ elements out of $NM$ are non-zero in each row and column of $\mh$. Let $\mathcal{I}_{d}$ and $\mathcal{J}_c$ denote the sets of non-zero positions in the $d^{th}$ row and $c^{th}$ column, respectively, then $|\mathcal{I}_{d}| = |\mathcal{J}_{c}| = {S}$.

Based on (\ref{vec_form}), we model the system as a sparsely connected factor graph with $NM$ variable nodes corresponding to $\vx$ and $NM$ observation nodes corresponding to $\vy$. In this factor graph, the observation node $ {y}_d$ is connected to the set of variable nodes $\{ {x}_e, e \in \mathcal{I}_d\}$. Similarly, the variable node $ {x}_c$ is connected to the set of variable nodes $\{ {y}_e, e \in \mathcal{J}_c\}$. 

\begin{figure}
\begin{minipage}[b]{0.5\columnwidth}
\centering
\includegraphics[scale=0.5]{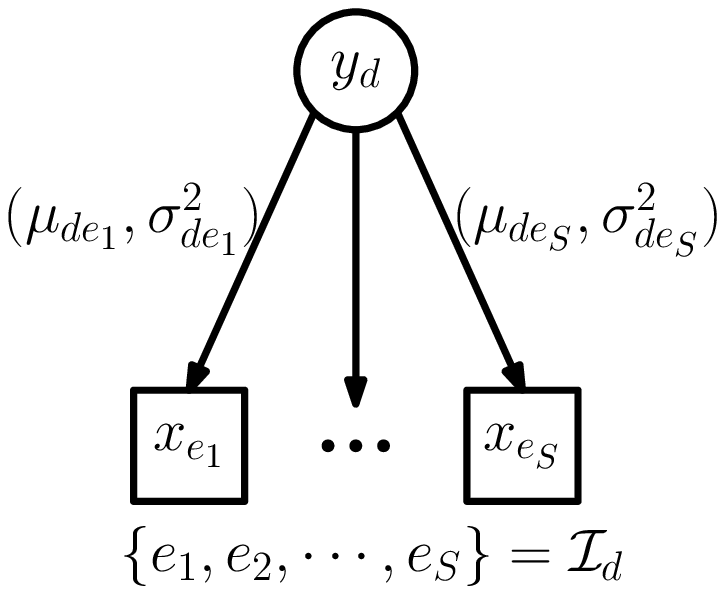}\\
{\footnotesize Observation node messages}
\end{minipage}%
\begin{minipage}[b]{0.5\columnwidth}
\centering
\includegraphics[scale=0.5]{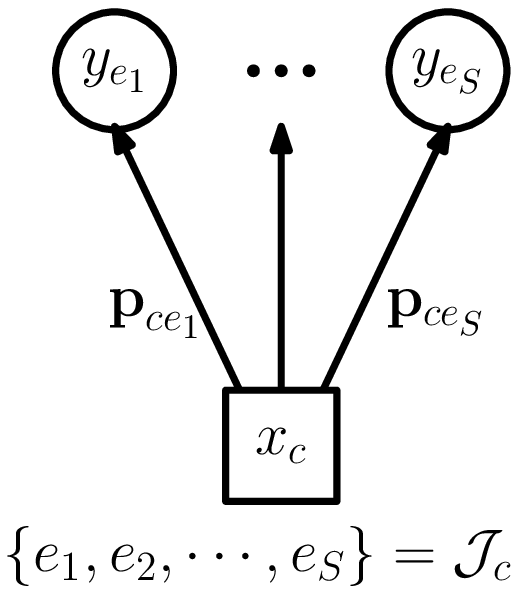}\\
{\footnotesize Variable node messages}
\end{minipage}
\caption{Messages in factor graph}
\label{mp_graph}
\vspace{-5mm}
\end{figure}

The joint maximum a posteriori probability (MAP) detection rule for estimating the transmitted information is given by
\begin{equation}
\widehat{{\bf x}} = \argmax_{ {{\bf x}} \in \mathbb{A}^{NM}} \, \Pr \left( \vx \Big| \vy,\mh \right),
\nonumber
\end{equation}
which has a complexity exponential in $NM$. Since the joint MAP detection is intractable even for very small values of $N$ and  $M$, we consider the symbol-by-symbol MAP detection rule for $0 \leq c \leq NM-1$,
\begin{align}
\widehat{x}_c & = \argmax_{ a_j \in \mathbb{A}} \, \Pr \left( x_c = a_j \Big| \vy,\mh \right)
 \\
& = \argmax_{ a_j \in \mathbb{A}} \, \frac{1}{|\mathbb{A}|} \Pr \left( \vy \Big| x_c = a_j,\mh \right)  \\
& \approx \argmax_{ a_j \in \mathbb{A}} \prod_{e \in \mathcal{J}_c} \Pr \left( y_e \Big| x_c = a_j,\mh \right).
\label{sym_map}
\end{align}
In  (\ref{sym_map}), we assume all the transmitted symbols $x_c \in \mathbb{A}$ are equally likely and the components of $\vy$ are approximately independent for a given $x_c$, due to the sparsity of $\mh$.
In order to solve the approximate symbol-by-symbol MAP detection in (\ref{sym_map}), we propose a MP detector which has a complexity linear in $NM$.
Similarly to \cite{mp_ga}, for each $y_e$, a variable $x_c$ is isolated from the other interference terms, which are then approximated as Gaussian noise with an easily computable mean and variance.

In MP, mean and variance of the interference terms are used as messages from observation nodes to variable nodes.
The message passed from a variable node ${x}_c$, for each  $c = \{0,\cdots,NM-1\}$, to the observation node ${y}_d$, for $d \in \mathcal{J}_c$, is the probability mass function (pmf)
${\bf p}_{cd} = \left\{ p_{cd}(a_j) |  a_j \in \mathbb{A} \right\}$ of the alphabet symbols in  $\mathbb{A}$.
Fig. \ref{mp_graph} shows the connections and the messages passed between the observation and variable nodes.

The MP algorithm operates as follows:\\
{\bf Step 1}: Initialize iteration index $i=0$ and ${\bf p}_{cd}^{(0)} = 1/|\mathbb{A}|$ for
 $c = \{0,\cdots,NM-1\}$ and $d \in \mathcal{J}_c$.   \\
{\bf Step 2}: Messages are passed from the observation nodes to the variable nodes.
The message passed from $ {y}_d$ to $ {x}_c$ is a Gaussian pdf 
which can be computed form 
\begin{align}
 {y}_d =  {x}_c  {H}_{d,c} + \underbrace{ \sum_{e \in \mathcal{I}_d, e \neq c}  {x}_e  {H}_{d,e} +  {z}_d}_{\zeta_{dc}},
 \label{gau_app}
\end{align}
where the interference-plus-noise term $\zeta_{dc}$ is approximated as Gaussian random variable with mean
\begin{align}
\mu_{dc}^{(i)} & = \mathbb{E} \left[ \zeta_{dc}  \right] = \sum_{e \in \mathcal{I}_d, e \neq c} \sum_{j=1}^{|\mathbb{A}|} p_{ed}^{(i)}(a_j) a_j  {H}_{d,e},
\label{mean_com}
\end{align}
and variance
\begin{small}
\begin{align}
& (\sigma^{(i)}_{dc})^2 = {\mbox {Var}} \left(\zeta_{dc}  \right) \label{var_com} \\
& = \! \! \! \! \! \sum_{e \in \mathcal{I}_d, e \neq c} \! \! \left( \sum_{j=1}^{|\mathbb{A}|} p_{ed}^{(i)}(a_j) |a_j|^2 | {H}_{d,e}|^2  \! - \left|\sum_{j=1}^{|\mathbb{A}|} p_{ed}^{(i)}(a_j) a_j  {H}_{d,e}\right|^2  \right) \! \! + \! \sigma^2 .
\nonumber
\end{align}
\end{small}
Further, we assume that transmitted symbols are i.i.d. and independent from the noise.\\
{\bf Step 3}: Messages are passed from variable nodes to the observation nodes. The new message from $ {x}_c$ to $ {y}_d$ contains the pmf vector ${\bf p}^{(i+1)}_{cd}$ with elements
\begin{align}
p_{cd}^{(i+1)}(a_j) & = \Delta \cdot  {p}_{cd}^{(i)}(a_j) + (1-\Delta) \cdot p_{cd}^{(i-1)} (a_j) ,
\label{prob_com}
\end{align}
where $\Delta \in (0,1]$ is the {\em damping factor} (\cite{damp}) to improve the convergence rate, and
\begin{align}
{p}_{cd}^{(i)}(a_j) \propto \prod_{e \in \mathcal{J}_c, e \neq d} \Pr \left( y_e \Big| x_c = a_j,\mh \right),
\nonumber
\end{align}
where
\[ \Pr \left( y_e \Big| x_c = a_j,\mh \right) \propto \exp \left( \frac{-\left| {y}_e - \mu_{ec}^{(i)} -  {H}_{e,c}a_j  \right|^2} {\sigma^{2,(i)}_{ec}} \right)
\]

 Note that this excludes the information of $ {y}_d$.\\
{\bf Step 4}: Repeat {\bf Step 2} and {\bf Step 3} until
\[
\displaystyle\max_{c,d,a_j} \left|{p}^{(i+1)}_{cd}(a_j) - {p}^{(i)}_{cd}(a_j)\right| < \epsilon
\]
or a maximum number of iterations is reached. \\
{\bf Step 5}: The final decisions about the transmitted symbols are obtained as
\begin{equation*}
\widehat{x}_c  = \argmax_{a_j \in \mathbb{A}} \, \, p_c(a_j), ~~ c \in \{0,\cdots,NM-1\}
\end{equation*}
where
\[p_c(a_j) = \prod_{e \in \mathcal{J}_c} \Pr \left( y_e \Big| x_c = a_j,\mh \right).
\]

{\em Complexity:}
The complexity of one iteration involves the computation of (\ref{mean_com}) and (\ref{var_com}),
where each computation has a complexity of the order $\mathcal{O} (NM S |\mathbb{A}|)$. Therefore, the overall complexity per symbol is $\mathcal{O} (n_{iter} S |\mathbb{A}|)$, where $n_{iter}$ is the number of iterations.
In simulations, we observed that the algorithm converges typically within 20 iterations. We  conclude that the sparsity of the delay-Doppler channel representation is a key factor in reducing the complexity of the decoder.
The memory requirement is dominated by the storage of $2NMS |\mathbb{A}|$ real values for ${\bf p}_{cd}^{(i)}$ and ${\bf p}_{cd}^{(i-1)}$. In addition, we have the massages $(\mu_{dc},\sigma^2_{dc})$, requiring $NMS$ complex values and $NMS$ real values, respectively.

\subsection{Application of MP detection algorithm for OFDM over delay--Doppler channels}
We now apply the above MP algorithm to OFDM to compensate the Doppler effects.

The OFDM system can be illustrated by the inner dashed box in Fig.~\ref{sys_fig}, i.e., the Time-Frequency domain.
Specifically, the Heisenberg Transform module is replaced by IFFT, cyclic prefix (CP) addition, serial-to-parallel and digital-to-analog conversion, and the Wigner Transform module is substituted with analog-to-digital, parallel-to-serial, CP removal and FFT operation. Also, as mentioned in {\it Remark 1}, for OFDM systems, $N$ is set to 1. 

In OFDM, the received signal $r(t)$ and noise $v(t)$ in (\ref{delay-doppler}) are sampled at $\frac{T}{M}$. Then, the frequency-domain signal after FFT operation is given by
\begin{align}
\mathbf{{y}} = \mathbf{W}\mathbf{{H}}_t\mathbf{W}^H\mathbf{{x}} + \mathbf{{v}}
\label{vec_form_ofdm}
\end{align}
where $(\cdot)^H$ denotes Hermitian transpose, $\mathbf{W}$ is $M$-point FFT matrix, and $\mathbf{{x}} \in \mathbb{A}^{M\times 1}$ is the transmitted information OFDM symbol.
The elements of time-domain channel matrix $\mathbf{{H}}_t$ are given as \cite{Zhao}
\[
{{H}_t}[p,q]\!=\!h_i\delta\!\left[\!\left[p-q-\frac{\tau_iM}{T}\right]_M\!\right]\!e^{j\frac{2\pi (q-1)\nu_i}{M}},\!\;p, q\!=1,\hdots,M.
\]
Using the frequency-domain channel matrix $\mathbf{{H}}\in\mathbb{C}^{M\times M} = \mathbf{W}\mathbf{{H}}_t\mathbf{W}^H$, we can re-write (\ref{vec_form_ofdm}) as:
\begin{align}
\mathbf{{y}} = \mathbf{{H}}\mathbf{{x}} + \mathbf{{v}}. \label{vec_form_ofdm1}
\end{align}
Since (\ref{vec_form_ofdm1}) has similar form as (\ref{vec_form}), the MP previously developed for OTFS can also be applied for OFDM symbol detection.
We note that  $\mh$ is {\em diagonally dominant} and the values of off-diagonal elements in each row decay as we move away from the diagonal entry. Hence, the $\mh$ matrix of OFDM is also sparse, which enables the use of the proposed low complexity MP detection algorithm.

{\remark} From (\ref{conv_eq}) and (\ref{vec_form_ofdm}), we can observe the effects of channel gain on the transmitted symbols are different in OTFS and OFDM. In OTFS, all the transmitted symbols experience the same channel gain (independent of $k$ and $l$), whereas in OFDM, the channel gains are distinct at different subcarriers because of the FFT operation on $\mh_t$.


\section{Illustrative Results}


In this section, we simulate the uncoded bit-error-rate (BER) performance of OTFS and OFDM over delay-Doppler channels. All relevant simulation parameters are given in Table \ref{tab_sp}.
For both OTFS and OFDM systems, Extended Vehicular A model\cite{LTE} is applied for the channel delay model, and the Doppler shift of the $i^{th}$ path is generated using
\[
\nu_i = \nu_{\mbox{max}}\cos(\theta_i),
\]
where $\theta_i\sim \mathcal{U}(0,\pi)$ is uniformly distributed.

\begin{table}[t]
\centering
  \begin{tabular}{ | r | p{2cm} | }
    \hline
    Parameter & Value  \\ \hline
    Carrier frequency  & 4 GHz  \\ \hline
    No. of subcarriers ($M$) & 512  \\ \hline
    No. of OTFS symbols ($N$) & 128 \\ \hline
    Subcarrier spacing & 15 KHz \\ \hline
    Cyclic prefix of OFDM & 2.6 $\mu$s\\ \hline
    Modulation alphabet & 4-QAM \\ \hline
    UE speed (km/h) & 30, 120, 500\\ \hline
    Channel estimation & Ideal \\ \hline
  \end{tabular}
  \vspace{2mm}
\caption{Simulation Parameters}
\vspace{-2mm}
\label{tab_sp}
\vspace{-5mm}
\end{table}

\begin{figure}
\centering
\includegraphics[width=3in,height=2.2in,clip=true,trim=1.5mm 1mm 1.4cm 7mm]{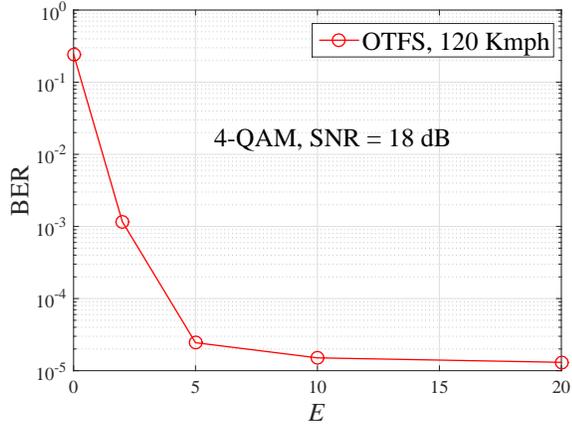}
\vspace{-2mm}
\caption{The BER performance of OTFS for different number of interference terms ($E$) with $4$-QAM.}
\label{sim1}
\end{figure}

Fig. \ref{sim1} shows the BER performance of OTFS system using the proposed MP detector for different number of interference terms ($E$) with $4$-QAM signaling at SNR $= 18$ dB over the delay-Doppler channel, where UE speed is $120$ km/h. Here, we consider $E_i = E, \forall i$. We can see that there is a significant performance improvement till $E = 10$ and saturation thereafter,  due to the IDI caused by the fractional Doppler. Fewer neighboring interference terms are sufficient to consider in MP (e.g. $5-10$).

\begin{figure}
\centering
\includegraphics[width=1.7in,height=1.8in,clip=true,trim=1.5mm 2mm 1.5cm 6mm]{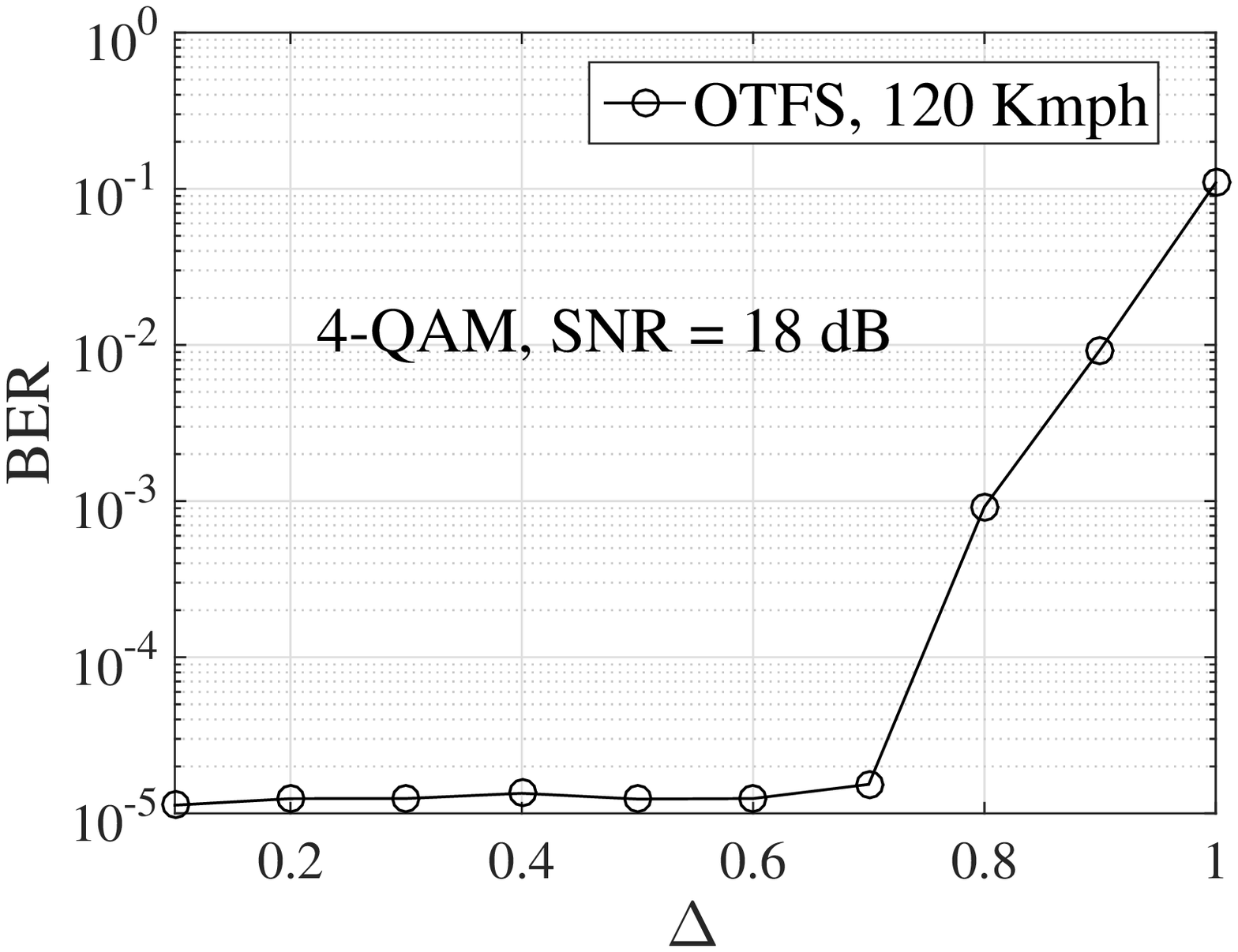}
\includegraphics[width=1.7in,height=1.8in,clip=true,trim=5mm 2mm 1.5cm 6mm]{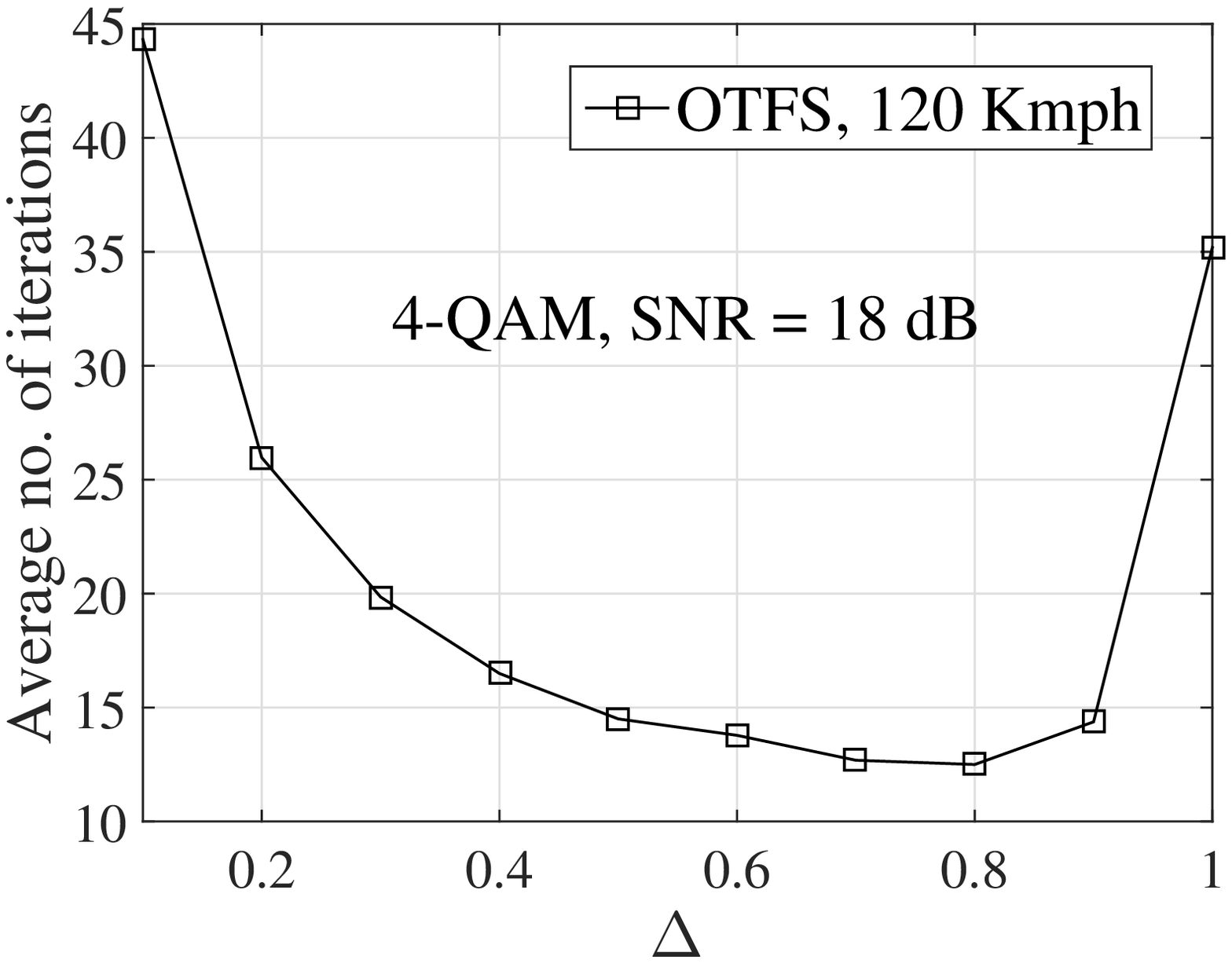}
\vspace{-2mm}
\caption{The variation of BER and average no. of iterations with $\Delta$.}
\label{delta_fig}
\vspace{-5mm}
\end{figure}

In Fig. \ref{delta_fig}, we illustrate the variation of BER and average number of iterations of OTFS using our MP detector over the delay-Doppler channel, where UE speed is $120$km/h. We adopt the damping factor $\Delta$ for $E = 10$. We consider $4$-QAM signaling and SNR $= 18$ dB. We observe that, when $\Delta \leq 0.7$, the BER of MP remains almost the same, but deteriorates thereafter.
Further, when $\Delta = 0.7$, MP converges with the least number of iterations. Hence, we choose $\Delta=0.7$ as the optimum damping factor.

\begin{figure}
\centering
\includegraphics[width=3in,height=2.2in,clip=true,trim=1.5mm 1mm 1.5cm 8mm]{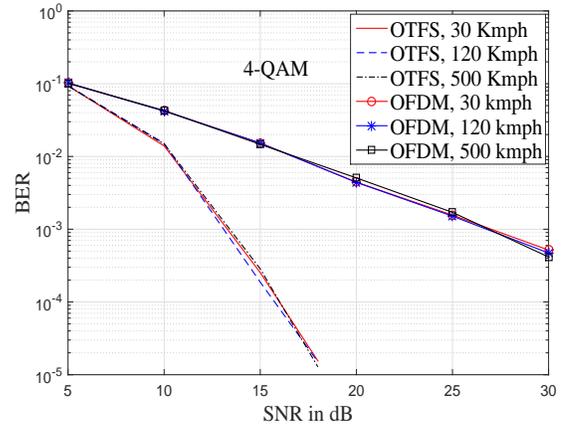}
\vspace{-2mm}
\caption{The BER performance comparison between OTFS and OFDM systems at different Doppler frequencies.}
\label{sim2}
\vspace{-5mm}
\end{figure}

In Fig. \ref{sim2}, we compare the BER performance of OTFS and OFDM systems using $4$-QAM signaling over the delay-Doppler channels of different Doppler frequencies (UE speeds of $30,120,500$km/h). We observe that OTFS outperforms OFDM by approximately $15$ dB at BER of $10^{-4}$, thanks to the constant channel gain over all transmitted symbols in OTFS, whereas in OFDM, the error performance is limited by the subcarrier with the lowest gain. Moreover, OTFS exhibits the same performance for different Doppler frequencies thanks to the IDI reduction provided by the MP detector and the assumption on $g_{tx}$ and $g_{rx}$. Similar behavior applies to OFDM, since the inter carrier interference (ICI) can be removed by the MP detector.

\section{Conclusion}
In this paper, we have analyzed the input--output  relation describing OTFS mod/demod in terms of sparse representation of the channel in the delay--Doppler domain. In particular, we have introduced the notion of inter Doppler interference caused by the fractional Doppler. We then proposed a linear complexity message passing (MP) detection algorithm which exploits the channel sparsity. Through simulations, we have shown that the effect of IDI can be mitigated by adapting the MP detection algorithm. We have also shown that OTFS has significant BER gains over OFDM under various channel conditions.

\section*{Acknowledgement}
This research work is support by the Australian Research Council under Discovery Project ARC DP160100528. Simulations were undertaken with the assistance of resources and services from the National Computational Infrastructure (NCI), which is supported by the Australian Government.


\begin{thebibliography}{1}
\bibitem{Hadani}
R. Hadani, S. Rakib, M. Tsatsanis, A. Monk, A. J. Goldsmith, A. F. Molisch, and R. Calderbank, ``Orthogonal time frequency space modulation,'' in {\it Proc. IEEE WCNC}, San Francisco, CA, USA, March 2017.


\bibitem{Jakes}
W.C Jakes, Jr., {\it Microwave Mobile Communications.} Wiley, New York, 1974.

\bibitem{LTE}
 E. LTE, ``Evolved universal terrestrial radio access (E-UTRA); base station (BS) radio transmission and reception (3GPP TS 36.104 version 8.6. 0 release 8), July 2009,'' ETSI TS, vol. 136, no. 104, p. V8.

\bibitem{wc_book}
D. N. C. Tse, P. Viswanath, {\it Fundamentals of wireless communications.} U.K., Cambridge: Cambridge Univ. Press, 2005.

\bibitem{Barbu}
  O. E. Barbu, C. N. Manch{\'o}n, C. Rom, and B. H. Fleury, ``Message-passing receiver for OFDM systems over highly delay-dispersive channels,'' {\em IEEE Trans. Wireless Commun.,} vol. 16, no. 3, pp. 1564-1578, March 2017.

\bibitem{Huang}
C. W. Huang, P. A. Ting, and C. C. Huang, ``A novel message passing based MIMO-OFDM data detector with a progressive parallel ICI canceller,'' {\em IEEE Trans. Wireless Commun.,} vol. 10, no. 4, pp. 1260-1268, April 2011.

\bibitem{damp}
M. Pretti, ``A message passing algorithm with damping,'' {\em J.~Stat.~Mech.: Theory and Experiment}, P11008, Nov. 2005.

\bibitem{mp_ga}
P. Som, T. Datta, N. Srinidhi, A. Chockalingam, and B. S. Rajan, ``Low-complexity detection in large-dimension MIMO-ISI channels using graphical models,'' {\em IEEE J.~Sel.~Topics in Signal Processing,} vol. 5, no. 8, pp. 1497-1511, December 2011.

\bibitem{Zhao}
Y. Zhao, and S. G. Haggman, ``Sensitivity to Doppler shift and carrier frequency errors in OFDM systems-the consequences and solutions,'' {\em in Proc. 46th IEEE Vehicular Technology Conf.}, Atlanta, GA, vol. 3, pp. 1564-1568, April 1996.

\bibitem{fut_pap}
P. Raviteja, K. T. Phan, Q. Jin, Y. Hong, and E. Viterbo, ``Interference cancellation for orthogonal time frequency space modulation'', {\em In preparation for submission to the IEEE Trans. Wireless Commun.}

\end{thebibliography}
\end{document}